\documentclass[aps,pra,twocolumn,showpacs]{revtex4}

\usepackage{graphicx}
\newcommand{\tr}{\mbox{Tr} }
\newcommand{\ket}[1]{\left | #1 \right \rangle}
\newcommand{\bra}[1]{\left \langle #1 \right |}
\newcommand{\proj}[1]{\ket{#1}\!\!\bra{#1}}

\newcommand{\hilbert}{{\cal H}}
\begin{document}
\title{Entropy exchange, coherent information and concurrence}
\author{Yang Xiang}
\email{njuxy@sina.com}
\author{Shi-Jie Xiong }

\affiliation{National Laboratory of Solid State Microstructures and
Department of Physics, Nanjing University, Nanjing 210093, China}
\date{\today}
\begin{abstract}
For a simple model we derive analytic expressions of entropy
exchange and coherent information, from which relations between them
and the concurrence are drawn. We find that in the quantum evolution
the entropy exchange exhibits behavior \textsl{opposite} to that of
the concurrence, whereas the coherent information shows features
very similar to those of the concurrence. The meaning of this result
for general systems is discussed.
\end{abstract}

\maketitle





Quantum entanglement is not only of interest in the interpretation
of the foundation of quantum mechanics, but also a resource useful
in quantum information processing and quantum computation. A measure
of the entanglement in a state of two qubits is the Wooters
concurrence \cite{wootters1}. For pure bipartite states one also
uses the Von Neumann entropy of the reduced density matrix for
either of the parties \cite{bennett1,popescu} to measure the
entanglement. For a simple example, which has been proposed by
Jordan \textsl{et al.} \cite{jordan}, we have shown that in the
quantum evolution a modified entanglement fidelity exhibits the
behavior similar to that of the concurrence \cite{xiang}. In this
paper we investigate other two quantities, the \textit{entropy
exchange} and the \textit{coherent information}
\cite{schumacher1,schumacher2}, by deriving analytic expressions for
the system proposed by Jordan \textsl{et al.}. From these we obtain
relations between them and the concurrence in the quantum evolution
and discuss the possibility of using them as measures of the
entanglement.

We begin with a brief discussion of noisy quantum processes and
their mathematical descriptions. $R$ and $Q$ are two quantum systems
and the joint system $RQ$ is initially prepared in a pure entangled
state $\ket{\Psi^{RQ}}$. The system $R$ is dynamically isolated and
has a zero internal Hamiltonian, while the system $Q$ undergoes some
evolution that possibly involves interaction with the environment.
The evolution of $Q$ might represent a transmission process via some
quantum channel for the quantum information in $Q$. In general, the
evolution of $Q$ can be represented by a quantum operator
$\varepsilon^{Q}$, which gives the mapping from the initial state to
the final state
\begin{eqnarray}
\rho^{Q'}=\varepsilon^{Q}(\rho^{Q}).
\end{eqnarray}
Here, $\rho^{Q}=\tr_{R}{\ket{\Psi^{RQ}}\bra{\Psi^{RQ}}}$ which
represents the initial state of system $Q$, and after the dynamical
process the final state of the system becomes $\rho^{Q'}$. In the
most general case, the map $\varepsilon^{Q}$ must be trace
preserving and is a linear positive map \cite{stinespring,kraus}, so
it can represent all unitary evolutions. The evolution may also
include unitary evolving interactions with an environment $E$.
Suppose that the environment is initially in state $\rho^{E}$. The
operator can be written as
\begin{eqnarray}
\varepsilon^{Q}(\rho^{Q})&=&\tr_{E}{U\left(\rho^{Q}\otimes\rho^{E}\right) U^{\dag}}\nonumber\\
&=&\tr_{E}{U\left( \rho^{Q}\otimes\sum_{i}{p_{i}|i\rangle\langle i|}\right) U^{\dag}}\nonumber\\
&=&\sum_{j}{E_{j}^{Q}\rho^{Q} E_{j}^{Q\dag}}, \label{operation}
\end{eqnarray}
where $\sum_{i}{p_{i}|i\rangle\langle i|}$ is the spectral
decomposition of $\rho^{E}$ with $\{ |i\rangle \}$ being a base in
the Hilbert space $\mathcal{H}_{E}$ of the environment $E$, and
$E^{Q}_{j}=\sum_{i}{\sqrt{p_{i}}\langle j|U|i\rangle}$. The final
state of $RQ$ is
\begin{eqnarray}
\rho^{RQ'}&= &\mathcal{I}^{R}\otimes\varepsilon^{Q}(\rho^{RQ})\nonumber\\
&=&\sum_{j}{(1^{R}\otimes E^{Q}_{j})\rho^{RQ}(1^{R}\otimes
E^{Q}_{j})^{\dag}}. \label{operation2}
\end{eqnarray}

The \textit{entropy exchange} $S_{e}$ is defined as
\cite{schumacher1}
\begin{eqnarray}
S_{e}=-\tr{\rho^{RQ'}\log{\rho^{RQ'}}}. \label{se1}
\end{eqnarray}
It is the von Neumann entropy of the final state of the joint system
$RQ$. Throughout this paper we use $\log$ to denote $\log_{2}$. The
intrinsic expression of the entropy exchange is given by
$S_{e}=S(W)$, $S(\rho)$ is the von Neumann entropy of density
operator $\rho$ and $W$ is a density operator with components (in an
orthonormal basis)
\begin{eqnarray}
W_{ij}=\tr{E^{Q}_{i}\rho^{Q} E^{Q\dag}_{j}}. \label{w1}
\end{eqnarray}
The entropy exchange is an intrinsic property of $Q$, depending only
on $\rho^{Q}$ and $\varepsilon^{Q}$. It is a measure of the
information exchanged between system $Q$ and the environment during
evolution $\varepsilon^{Q}$. If there is no interaction between
system $Q$ and the environment, i.e., $\varepsilon^{Q}$ is a unitary
operator, then after the dynamical process the final state of the
joint system $RQ$ is still a pure state. This means that in this
case the entropy exchange equals zero. A complete discussion of the
entropy exchange can be seen in Refs.
\cite{schumacher1,schumacher2,barnum}. In Ref. \cite{xiang} we show
that the modified entanglement fidelity (MEF) admirably reflects the
entanglement preservation. The MEF is defined as
\begin{eqnarray}
F_{e}=\max_{U^{Q}}\bra{\Psi^{RQ}}(1^{R}\otimes
U^{Q})\rho^{RQ'}(1^{R}\otimes U^{Q})^{\dag}\ket{\Psi^{RQ}},
\label{f}
\end{eqnarray}
where $U^{Q}$ is a unitary transformation acting on $Q$. A
connection between the MEF and the entropy exchange is given by the
\textit{quantum Fano inequality} \cite{schumacher1},
\begin{eqnarray}
h(F_{e})+(1-F_{e})\log(d^{2}-1)\geq S_{e}, \label{fano}
\end{eqnarray}
where $h(\rho)=-\rho\log\rho-(1-\rho)\log(1-\rho)$ and $d$ is the
number of the complex dimensions of the Hilbert space $\hilbert$
describing system $Q$. We can easily find that when $F_{e}=1$, i.e.,
the entanglement is admirably preserved in the evolution process,
$S_{e}$ equals zero. This implies that there may be some connection
between the entropy exchange and the measures of the entanglement,
e.g., the concurrence.

Another important intrinsic quantity is the \textit{coherent
information} $I_{e}$. It is defined as \cite{schumacher2}
\begin{eqnarray}
I_{e}=S(\rho^{Q'})-S(\rho^{RQ'})=S(\rho^{Q'})-S_{e}.
\label{i}
\end{eqnarray}
For classical systems, $I_{e}$ can never be positive since the
entropy of the joint system $RQ$ can never be less than the entropy
of the subsystem $Q$. But, for quantum systems, $I_{e}$ may take
positive value. For example, if $\rho^{RQ'}$ is an entangled pure
state, then $\rho^{Q'}$ is a mixed state, implying that
$S(\rho^{Q'})>0$ and $S(\rho^{RQ'})=0$. Thus, $I_{e}$ can be
regarded as a measure of the ``nonclassicity'' of the final joint
state $\rho^{RQ'}$, or, in other words, the degree of the quantum
entanglement retained by $R$ and $Q$.

Now we discuss a simple example introduced by Jordan \textsl{et al.}
in their work \cite{jordan}. For this example we can give analytic
expressions of the entropy exchange and the coherent information
from which the relations between them and the concurrence can be
obtained. We consider two entangled qubits, $A$ and $B$, and suppose
that qubit $A$ interacts with a control qubit $C$. Then $A$, $B$ and
$C$ respectively correspond to systems $Q$, $R$ and environment $E$
discussed above. We suppose that the initial states of the three
qubits are
\begin{eqnarray}
\Lambda=\rho^{AB}\otimes\frac{1}{2} 1_{c} \label{w},
\end{eqnarray}
where $\rho^{AB}$ is the initial state of joint system $A$ and $B$.
According to the Schmidt theorem \cite{schmidt}, a pure state of two
$\frac{1}{2}$-spins can be decomposed as
\begin{eqnarray}
\ket{\Psi}&=&e^{-i\varphi/2}\cos(\frac{\theta}{2})\ket{{\bf
n}}_{A}\ket{{\bf
m}}_{B}\nonumber\\
&~~~~&+e^{i\varphi/2}\sin(\frac{\theta}{2})\ket{{\bf
-n}}_{A}\ket{{\bf -m}}_{B}, \label{www}
\end{eqnarray}
where ${\bf n}$ and ${\bf m}$ are two points on the Poincar\'{e}
sphere, and the subscript specifies the related qubit $A$ or $B$.
The ``angle'' $\theta$ in Eq. (\ref{www}) determines the degree of
entanglement in the state. The angle satisfies $0\leq\theta\leq\pi$,
$\theta=0$ and $\theta=\pi$ correspond to the product states, and
the maximal entanglement is obtained for $\theta=\frac{\pi}{2}$.
Without losing generality we set the initial state of the joint
system $A$ and $B$ as
\begin{eqnarray}
\ket{\psi}=\cos\left(\frac{\theta}{2}\right)\ket{+z}_{A}\ket{-z}_{B}
+\sin\left(\frac{\theta}{2}\right)\ket{-z}_{A}\ket{+z}_{B},
\label{initial state}
\end{eqnarray}
where $\ket{\pm z}$ represent the eigenstates of $\sigma_{z}$ with
eigenvalues $\pm 1$. So the density matrix
$\rho^{AB}=\ket{\psi}\bra{\psi}$. For simplicity, we omit the
``azimuthal angle" $\varphi$, as factor $e^{\pm i\varphi/2}$ can be
absorbed in eigenstates $\ket{\pm z}$ and has no effect on the final
results.

We suggest an interaction between qubits $A$ and $C$ described by
the unitary transformation
\begin{eqnarray}
U=e^{-i t H} \label{u},
\end{eqnarray}
where
\begin{eqnarray}
H=\frac{\lambda\sigma^{A}_{z}}{2}(\ket{\alpha}\bra{\alpha}-\ket{\beta}\bra{\beta})
\label{h}
\end{eqnarray}
with $\lambda$ being the strength of the interaction, and
$\ket{\alpha}$ and $\ket{\beta}$ are two orthonormal vectors for
system $C$. Then the changing density matrix for joint system of $A$
and $B$ can be calculated as
\begin{widetext}
\begin{eqnarray}
\rho^{AB'}&=&\tr_{c}{\left[(U\otimes1^{B})\Lambda(U\otimes1^{B})^{\dag}\right]}
=\tr_{c}{\left[(U\otimes1^{B})(\ket{\psi}\bra{\psi}\otimes\frac{1}{2}1_{c})(U\otimes1^{B})^{\dag}\right]}\nonumber\\
&=&\left(
\begin{array}{c}
0~~~~~~~~0~~~~~~~~~~~~~~~~~~~~~~~~~~~~~~~0~~~~~~~~~~~~~~~~~~~~~~~~~~~~~~~~~~~~~~~~~~~0\\
0~~~~~~~\cos^{2}\left(\frac{\theta}{2}\right)~~~~~~~~~~~~~~~~~~~~~\cos(\lambda
t)\cos\left(\frac{\theta}{2}\right)\sin\left(\frac{\theta}{2}\right)~~~~~~~~~~~~~~~~0\\
0~~~~~~~\cos(\lambda
t)\cos\left(\frac{\theta}{2}\right)\sin\left(\frac{\theta}{2}\right)~~~~\sin^{2}\left(\frac{\theta}{2}\right)
~~~~~~~~~~~~~~~~~~~~~~~~~~~~~~~~~~0\\
0~~~~~~~~0~~~~~~~~~~~~~~~~~~~~~~~~~~~~~~~0~~~~~~~~~~~~~~~~~~~~~~~~~~~~~~~~~~~~~~~~~~~0
\end{array}\right). \label{dm}
\end{eqnarray}
\end{widetext}

The changing density matrix $\rho^{AB'}$ usually represents a mixed
state. In order to quantify the entanglement we adopt the Wootters
concurrence \cite{wootters1} defined as
\begin{eqnarray}
C(\rho)\equiv\max[0,\sqrt{\lambda_{1}}-\sqrt{\lambda_{2}}-
\sqrt{\lambda_{3}}-\sqrt{\lambda_{4}}],
\end{eqnarray}
where $\rho$ is the density matrix representing the investigated
state of the joint system of $A$ and $B$, $\lambda_{1}$,
$\lambda_{2}$, $\lambda_{3}$, and $\lambda_{4}$ are the eigenvalues
of $\rho\sigma^{A}_{2}\sigma^{B}_{2}\rho^{\ast}
\sigma^{A}_{2}\sigma^{B}_{2}$ in the decreasing order, and
$\rho^{\ast}$ is the complex conjugation of $\rho$.
\begin{figure}[htb]
  \centering
  \begin{minipage}[c]{0.5\columnwidth}
    \centering
    \includegraphics[width=1\columnwidth,
height=0.8\columnwidth]{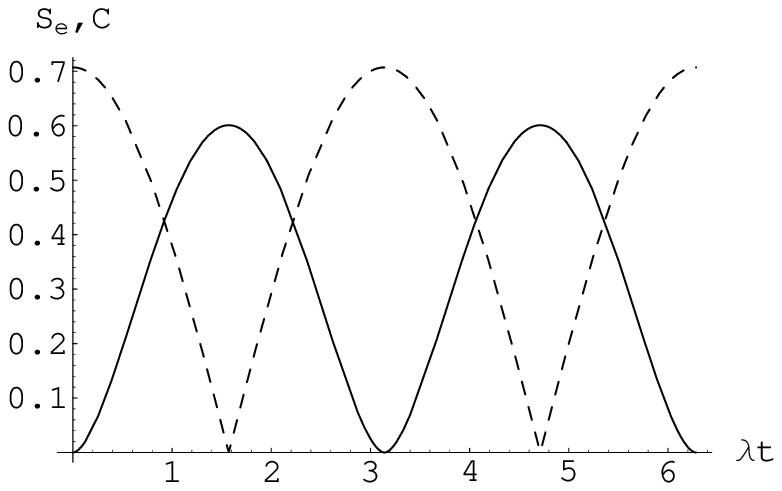}\\
    {(a)$\;\theta=\pi/4$}
  \end{minipage}%
  \begin{minipage}[c]{0.5\columnwidth}
    \centering
    \includegraphics[width=1\columnwidth,
height=0.8\columnwidth]{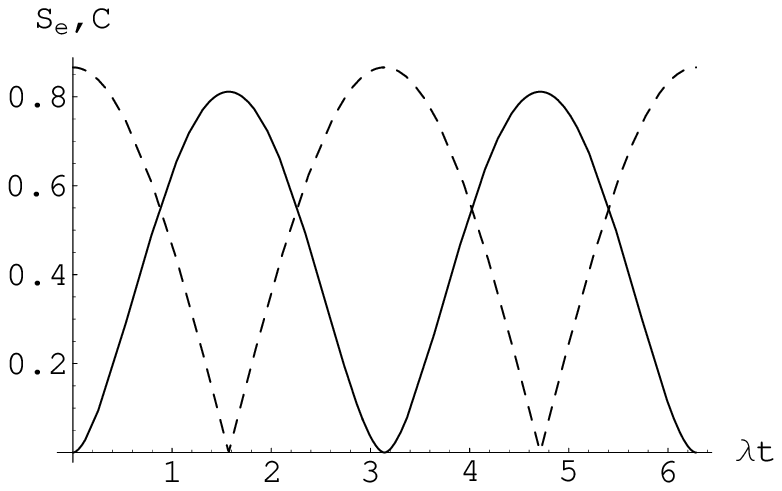}\\
    {(b)$\;\theta=\pi/3$}
  \end{minipage}
  \vspace*{0.5cm}
  \begin{minipage}[c]{0.5\columnwidth}
    \centering
    \includegraphics[width=1\columnwidth,
height=0.8\columnwidth]{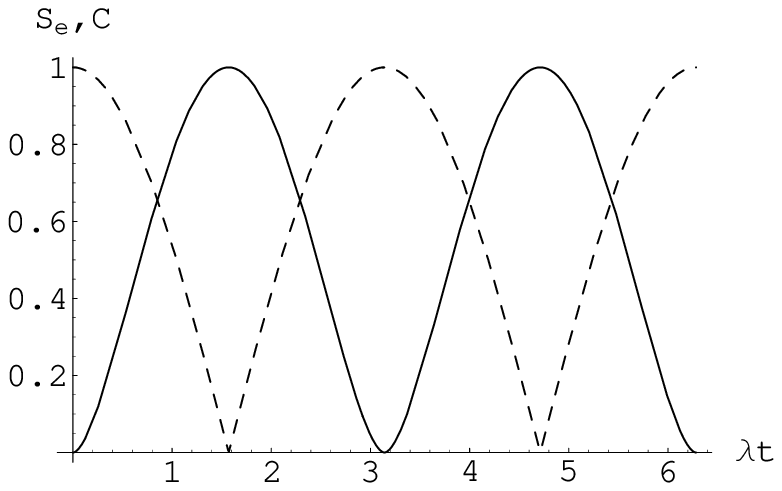}\\
    {(c)$\;\theta=\pi/2$}
  \end{minipage}%
  \begin{minipage}[c]{0.5\columnwidth}
    \centering
    \includegraphics[width=1\columnwidth,
height=0.8\columnwidth]{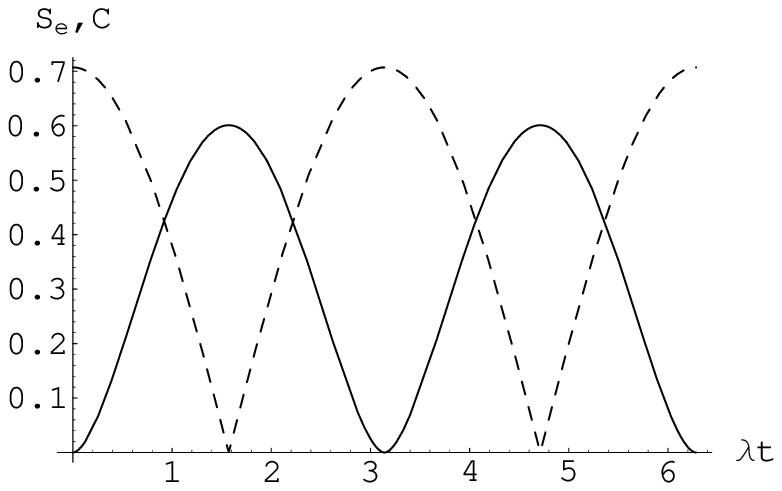}\\
    {(d)$\;\theta=3\pi/4$}
  \end{minipage}
\caption{Evolutions of the entropy exchange $S_{e}$ (solid line) and
the concurrence $C$ (dashed line). We take $\hbar$=1 so that
$\lambda t$ is dimensionless.}
 \label{fig1}
\end{figure}
From Eq. (\ref{dm}) we can obtain
\begin{eqnarray}
C(\rho^{AB'})=\sin\theta|\cos{\lambda t}|.
\end{eqnarray}
It is found that at time $\lambda t=\frac{\pi}{2}$, the state
$\rho^{AB'}$ is changed from the initial entangled state at $t=0$ to
a separable state, while at time $\lambda t=\pi$ the state
$\rho^{AB'}$ returns to the state with the initial entanglement. We
can also find that if the initial state is a product state, i.e.,
$\theta=0$ or $\theta=\pi$, the concurrence always equals zero.

Using Eqs. (\ref{operation}), (\ref{w}), (\ref{u}), and (\ref{h}),
we obtain the quantum operation on qubit $A$,
\begin{eqnarray}
\varepsilon^{A}(\rho^{A})&=&\tr_{C}{U(\rho^{A}\otimes\rho^{C})U^{\dag}}\nonumber\\
&=&\tr_{C}{U\left(\rho^{A}\otimes(\frac{1}{2}(\proj{\alpha}+\proj{\beta}))\right)U^{\dag}}\nonumber\\
&=&\frac{1}{2}e^{-i \sigma^{A}_{3}\left(\frac{\lambda
t}{2}\right)}\rho^{A}e^{+i \sigma^{A}_{3}\left(\frac{\lambda
t}{2}\right)}\nonumber\\
&~~~~&+\frac{1}{2}e^{+i \sigma^{A}_{3}\left(\frac{\lambda
t}{2}\right)}\rho^{A}e^{-i \sigma^{A}_{3}\left(\frac{\lambda
t}{2}\right)}. \label{expression}
\end{eqnarray}
So $E^{A}_{\alpha}=\frac{1}{\sqrt{2}}e^{-i
\sigma^{A}_{3}\left(\frac{\lambda t}{2}\right)}$ and
$E^{A}_{\beta}=\frac{1}{\sqrt{2}}e^{+i
\sigma^{A}_{3}\left(\frac{\lambda t}{2}\right)}$. Substituting them
into Eq. (\ref{w1}) and noting that $\rho^{A} \equiv
\tr_{B}{(\rho^{AB})}=\left(
\begin{array}{c}
\cos^{2}\left(\frac{\theta}{2}\right)~0~~~~~~~~~\\
0~~~~~~~~~~\sin^{2}\left(\frac{\theta}{2}\right)
\end{array}\right)$, we can get the density matrix $W$ in Eq. (\ref{w1}) as
\begin{widetext}
\begin{eqnarray}
W=\frac{1}{2}\left(\begin{array}{c}
1~~~~~~~~~~~~~~~~~~~~~~~~~~~~~~~~\cos(\lambda t)-i\sin(\lambda t)\cos(\theta)\\
\cos(\lambda t)+i\sin(\lambda
t)\cos(\theta)~~1~~~~~~~~~~~~~~~~~~~~~~~~~~~~~~
\end{array}\right). \label{w2}
\end{eqnarray}
Thus, we have
\begin{eqnarray}
S_{e}=&-&\frac{1}{2}\left[1-\sqrt{\cos^{2}(\lambda
t)+\cos^{2}(\theta)\sin^{2}(\lambda
t)}\right]\log\left\{\frac{1}{2}\left[1-\sqrt{\cos^{2}(\lambda
t)+\cos^{2}(\theta)\sin^{2}(\lambda
t)}\right]\right\}\nonumber\\
&-&\frac{1}{2}\left[1+\sqrt{\cos^{2}(\lambda
t)+\cos^{2}(\theta)\sin^{2}(\lambda
t)}\right]\log\left\{\frac{1}{2}\left[1+\sqrt{\cos^{2}(\lambda
t)+\cos^{2}(\theta)\sin^{2}(\lambda t)}\right]\right\}. \label{se}
\end{eqnarray}
\end{widetext}
We can also directly calculate $S_{e}$ by using Eq. (\ref{dm}) and
Eq. (\ref{se1}). From Eq. (\ref{se}), we find that if the initial
state is a product one, i.e., $\theta=0$ or $\theta=\pi$, the
entropy exchange always equals zero.

The evolutions of the entropy exchange $S_{e}$ and the concurrence
$C(\rho^{AB'})$ are depicted in Fig. \ref{fig1}. We can see that the
entropy exchange exhibits the behavior \textsl{opposite} to that of
the concurrence in the quantum evolution. During the evolution
$\varepsilon^{A}$, the more information exchanged between $A$ and
``environment'' $C$, the more entanglement between $A$ and $B$ is
lost.

\begin{figure}[htb]
  \centering
  \begin{minipage}[c]{0.5\columnwidth}
    \centering
    \includegraphics[width=1\columnwidth,
height=0.8\columnwidth]{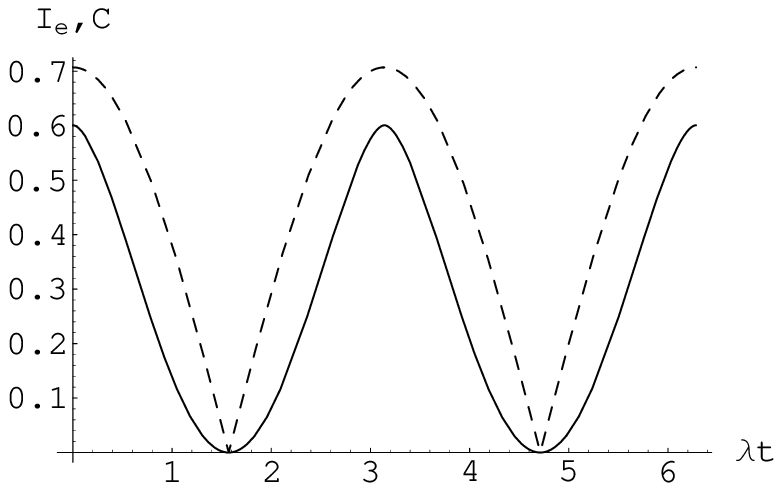}\\
    {(a)$\;\theta=\pi/4$}
  \end{minipage}%
  \begin{minipage}[c]{0.5\columnwidth}
    \centering
    \includegraphics[width=1\columnwidth,
height=0.8\columnwidth]{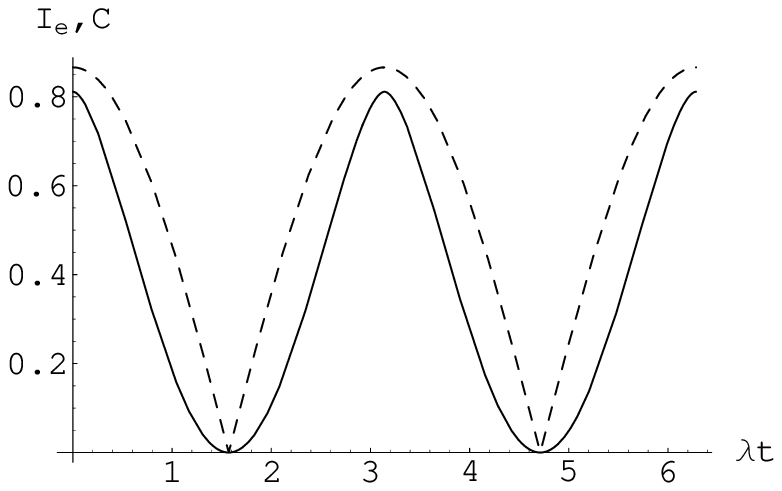}\\
    {(b)$\;\theta=\pi/3$}
  \end{minipage}
  \vspace*{0.5cm}
  \begin{minipage}[c]{0.5\columnwidth}
    \centering
    \includegraphics[width=1\columnwidth,
height=0.8\columnwidth]{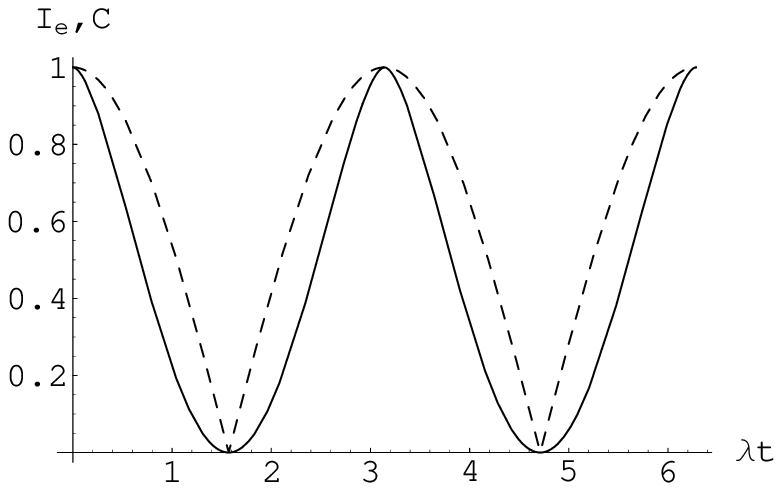}\\
    {(c)$\;\theta=\pi/2$}
  \end{minipage}%
  \begin{minipage}[c]{0.5\columnwidth}
    \centering
    \includegraphics[width=1\columnwidth,
height=0.8\columnwidth]{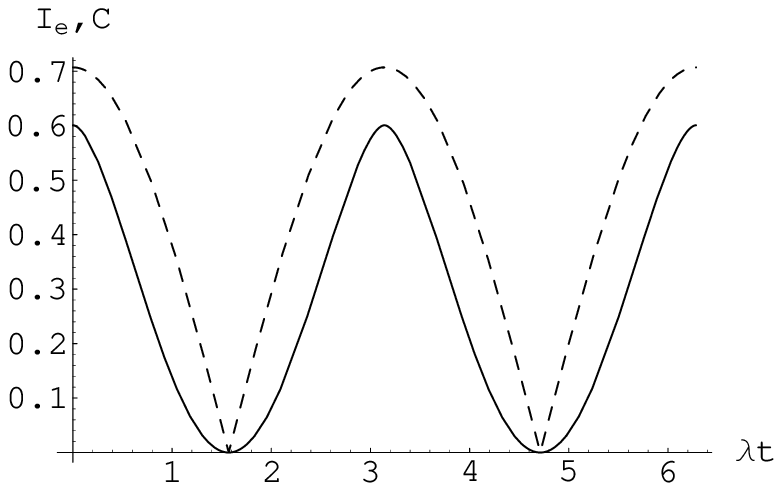}\\
    {(d)$\;\theta=3\pi/4$}
  \end{minipage}
\caption{Evolutions of the coherent information $I_{e}$ (solid line)
and the concurrence $C$ (dashed line). }
 \label{fig2}
\end{figure}

Now we discuss the coherent information $I_{e}$. By using Eq.
(\ref{dm}) we find that $\rho^{A'}= \tr_{B}{(\rho^{AB'})}=\left(
\begin{array}{c}
\cos^{2}\left(\frac{\theta}{2}\right)~0~~~~~~~~~\\
0~~~~~~~~~~\sin^{2}\left(\frac{\theta}{2}\right)
\end{array}\right)$, so from Eq. (\ref{i}) we can obtain $I_{e}$.
The evolutions of the coherent information $I_{e}$ and the
concurrence $C(\rho^{AB'})$ are depicted in Fig. \ref{fig2}. It can
be seen that $I_{e}$ exhibits the behavior very similar to that of
the concurrence in the quantum evolution, although their values are
not exactly consistent with each other at all moments.

The negative correlation between the entropy exchange and the
concurrence is a striking feature of the quantities. In this example
the range of $S_{e}$ is in $(0,1)$, where the left-hand side of the
Fano inequality (\ref{fano}) is a monotonic decreasing function of
the MEF $F_{e}$. This means that for a quantum operation with a
given $S_{e}$ there should be an upper bound of $F_{e}$. In Ref.
\cite{xiang} we have shown that $F_{e}$ exhibits the behavior
similar to that of the concurrence in the quantum evolution. So for
a quantum operation giving $S_{e}$ there is also an upper bound of
the concurrence. This only gives a rough correlation between the
concurrence and the entropy exchange. Owing to the simplicity of the
example we are able to obtain the analytical expressions of $F_{e}$
and $S_{e}$ and to illustrate the definite negative correlation
between them. In this sample the effect of the environment is
represented by only a single qubit ($C$). In Eq. (\ref{w}), we
assumed the initial state of qubit $C$ is a mixed one. For a more
complicated environment we may introduce an extra system, e.g., a
qubit $D$, in addition to qubit $C$. If this new qubit purifies the
initial state, then all the results obtained above are retained. In
fact, we can regard the system $AB$ and the environment $CD$ as a
new joint system and assume that the initial state of this joint
system is a pure state. After a quantum operation which represents
the interaction between subsystems $A$ and $C$, the final state of
the joint system must be a pure state too, so the entropy exchange
can also be seen as a measure of the entanglement between $AB$ and
$CD$. The information exchange between $A$ and $C$, which is the
result of the interaction, will cause the entanglement between $AB$
and $CD$, and decrease the entanglement between $A$ and $B$. This
results in the opposite behavior between the entropy exchange and
the concurrence in the quantum evolution. This also means that the
entanglement between $A$ and $B$ and the entanglement between $AB$
and $CD$ have negative correlation. Thus, the conclusion about the
negative correlation between the entropy exchange and the
concurrence may be extended to the case of environment having two
qubits but the initial state being purified. In other cases where
the environment includes more qubits but there are no correlations
between them, the effect of every qubit in the environment should be
similar to that of the single qubit investigated in this example.

The coherent information $I_{e}$ depends only on $\rho^{Q}$ and
$\varepsilon^{Q}$ and is used to measure the amount of quantum
information conveyed in the noisy channel. We imagine that the
initial state of system $Q$ is $\rho^{Q}$ which arises from the
entanglement between $Q$ and a reference system $R$. Now the goal of
Alice in $Q$ is to send the initial state $\rho^{Q}$ to Bob via a
quantum channel (which can be described by $\varepsilon^{Q}$) and
establish an entanglement between reference system $R$ and Bob's
output system $Q'$. The coherent information $I_{e}$ can express the
degree of achieving this aim. In above investigation we just model
the time evolution of quantum information transmitted via a noisy
quantum channel described by the interaction with a control qubit.
By comparing the coherent information $I_{e}$ and the concurrence,
we find that $I_{e}$ is indeed a good measure for the capacity of a
quantum channel to transmit the entanglement.

In summary, for an example modeling interaction with environment we
derive the analytic expressions of the entropy exchange and the
coherent information. From these we find that both the entropy
exchange and the coherent information have profound correlations
with the measure of the entanglement.


\vskip 0.5 cm

{\it Acknowledgments} This work was supported by the State Key
Programs for Basic Research of China (Grants No. 2005CB623605 and
No. 2006CB921803), and by National Foundation of Natural Science in
China Grant No. 10474033 and No. 60676056.


\end{document}